\documentclass[twocolumn,superscriptaddress,preprintnumbers,amsmath,amssymb,prl]{revtex4}
\usepackage{graphicx}
\usepackage{dcolumn,xcolor}
\usepackage[normalem]{ulem}
\usepackage{amsmath} 
\usepackage{amssymb}
\usepackage{amsfonts}
\usepackage{bm}
\usepackage[latin1]{inputenc}
\usepackage[dvips]{epsfig}
\usepackage{hyperref}
\usepackage{color}

\definecolor{fig1_1}{rgb}{1.000000000000000 ,  0.781200000000000  , 0.497500000000000}
\definecolor{fig1_2}{rgb}{1.000000000000000 ,  0.694400000000000  , 0.442222222222222}
\definecolor{fig1_3}{rgb}{0.972222222222222 ,  0.607600000000000  , 0.386944444444444}
\definecolor{fig1_4}{rgb}{0.833333333333333 ,  0.520800000000000  , 0.331666666666667}
\definecolor{fig1_5}{rgb}{0.694444444444444   0.434000000000000   0.276388888888889}
\definecolor{fig1_6}{rgb}{0.555555555555556   0.347200000000000   0.221111111111111}
\definecolor{fig1_7}{rgb}{0.416666666666667 ,  0.260400000000000  , 0.165833333333333}
\definecolor{fig1_8}{rgb}{0.277777777777778 ,  0.173600000000000  , 0.110555555555556}
\definecolor{fig1_9}{rgb}{0.138888888888889 ,  0.086800000000000  , 0.055277777777778}
\definecolor{fig2a_1}{rgb}{0,0.25,0.875}
\definecolor{fig2a_2}{rgb}{0,0.5,0.75}
\definecolor{fig2a_3}{rgb}{0,0.75,0.625}
\definecolor{fig2a_4}{rgb}{0,1,0.5}
\definecolor{fig2b_1}{rgb}{0.5,0.31248,0.199}
\definecolor{fig2b_2}{rgb}{0.75,0.46872,0.2985}
\definecolor{fig2b_3}{rgb}{1,0.62496,0.398}
\definecolor{fig2b_4}{rgb}{1,0.7812,0.4975}
\definecolor{fig4_1}{rgb}{1,0.7812,0.4975}
\definecolor{fig4_2}{rgb}{1,0.651,0.41458}
\definecolor{fig4_3}{rgb}{0.83333,0.5208,0.33167}
\definecolor{fig4_4}{rgb}{0.625,0.3906,0.24875}
\definecolor{fig4_5}{rgb}{0.41667,0.2604,0.16583}
\definecolor{sfig2a_1}{rgb}{1,0,0}
\definecolor{sfig2b_1}{rgb}{0,0,1}
\definecolor{sfig2a_2}{rgb}{1,0.2,0}
\definecolor{sfig2b_2}{rgb}{0,0.2,0.9}
\definecolor{sfig2a_3}{rgb}{1,0.4,0}
\definecolor{sfig2b_3}{rgb}{0,0.4,0.8}
\definecolor{sfig2a_4}{rgb}{1,0.6,0}
\definecolor{sfig2b_4}{rgb}{0,0.6,0.7}
\definecolor{sfig2a_5}{rgb}{1,0.8,0}
\definecolor{sfig2b_5}{rgb}{0,0.8,0.6}
\definecolor{fig2a_1}{rgb}{0,0,1}
\definecolor{fig2a_2}{rgb}{0,0.2,0.9}
\definecolor{fig2a_3}{rgb}{0,0.4,0.8}
\definecolor{fig2a_4}{rgb}{0,0.6,0.7}
\definecolor{fig2a_5}{rgb}{0,0.8,0.6}

\begin{document}

\title{Unified theoretical and experimental view on transient shear banding}

 \author{Roberto Benzi}
 \affiliation{Dipartimento di Fisica, Universit\`a di Roma ``Tor Vergata" and INFN, Via della Ricerca Scientifica, 1-00133 Roma, Italy}
 \author{Thibaut Divoux}
  \affiliation{MultiScale Material Science for Energy and Environment, UMI 3466, CNRS-MIT, 77 Massachusetts Avenue, Cambridge, Massachusetts 02139, USA}
  \affiliation{Department of Civil and Environmental Engineering, Massachusetts Institute of Technology, Cambridge, MA 02139}
 \author{Catherine Barentin}
  \affiliation{Universit\'e de Lyon, Universit\'e Claude Bernard Lyon 1, CNRS, Institut Lumi\`ere Mati\`ere, F-69622 Villeurbanne, France}
 \author{S\'ebastien Manneville}
  \affiliation{MultiScale Material Science for Energy and Environment, UMI 3466, CNRS-MIT, 77 Massachusetts Avenue, Cambridge, Massachusetts 02139, USA}
\affiliation{Univ Lyon, Ens de Lyon, Univ Claude Bernard, CNRS, Laboratoire de Physique, F-69342 Lyon, France}
\author{Mauro Sbragaglia}
\affiliation{Dipartimento di Fisica, Universit\`a di Roma ``Tor Vergata" and INFN, Via della Ricerca Scientifica, 1-00133 Roma, Italy}
\author{Federico Toschi}
\affiliation{Department of Applied Physics, Eindhoven University of Technology, P.O. Box 513, 5600 MB Eindhoven, The Netherlands}

\date{\today}

\begin{abstract}
Dense emulsions, colloidal gels, microgels, and foams all display a solid-like behavior at rest characterized by a yield stress, above which the material flows like a liquid. Such a fluidization transition often consists of long-lasting transient flows that involve shear-banded velocity profiles. The characteristic time for full fluidization, $\tau_\text{f}$, has been reported to decay as a power-law of the shear rate $\dot \gamma$ and of the shear stress $\sigma$ with respective exponents $\alpha$ and $\beta$. Strikingly, the ratio of these exponents was empirically observed to coincide with the exponent of the Herschel-Bulkley law that describes the steady-state flow behavior of these complex fluids. Here we introduce a continuum model, based on the minimization of a {``free energy''}, that captures quantitatively all the salient features associated with such \textit{transient} shear-banding. More generally, our results provide a unified theoretical framework for describing the yielding transition and the steady-state flow properties of yield stress fluids. 
\end{abstract}

\maketitle

\textit{Introduction.-} Amorphous soft materials, such as dense emulsions, foams and microgels, display solid-like properties at rest, while they flow like liquids for large enough stresses \cite{Barnes:1999,Balmforth:2014,Coussot:2015,Bonn:2017}. These yield stress fluids are characterized by a steady-state flow behavior that is well described by the Herschel-Bulkley (HB) model, where the shear stress $\sigma$ is linked to the shear rate $\dot \gamma$ through $\sigma=\sigma_\text{c}+A\dot \gamma^n$, with $\sigma_\text{c}$ the yield stress of the fluid, $A$ the consistency index and $n$ a phenomenological exponent that ranges between 0.3 and 0.7, and is often equal to $1/2$ \cite{Herschel:1926,Barnes:2001,Katgert:2008,Cohen:2014}. However, steady-state flow is never reached instantly and the yielding transition may involve transient regimes much longer than the natural timescale $\dot \gamma^{-1}$ \cite{Sprakel:2011,Siebenburger:2012a,Grenard:2014,Fielding:2014,Divoux:2016,Bonn:2017}.

\begin{figure}[t!]
\centering
\includegraphics[width=0.8\linewidth]{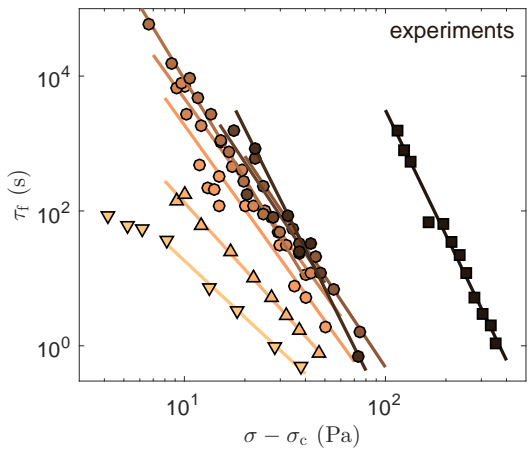}
\caption{(color online) Stress-induced fluidization time $\tau_\text{f}$ vs reduced shear stress $\sigma-\sigma_\text{c}$ for carbopol microgels at various weight concentrations: 0.5\% (\textcolor{fig1_1}{$\blacktriangledown$}), 0.7\% (\textcolor{fig1_2}{$\blacktriangle$}), 1\% (\textcolor{fig1_5}{$\bullet$}) and 3\% (\textcolor{fig1_9}{\tiny $\blacksquare$}). Solid lines correspond to the best power-law fits of the various data sets $\tau_\text{f}\sim(\sigma-\sigma_\text{c})^{-\beta}$ with exponent $\beta$ ranging from 2.8 to 6.2. Experimental conditions are listed in Supplemental Table~S1 together with values of $\sigma_\text{c}$ and $\beta$.
\label{fig1}}
\end{figure}

As demonstrated experimentally in Refs. \cite{Divoux:2010,Divoux:2011b,Divoux:2012}, long-lasting heterogeneous flows develop from the initial solid-like state, involving shear-banded velocity profiles before reaching a homogeneous steady-state flow. Depending on the imposed variable, $\dot \gamma$ or $\sigma$, the characteristic time $\tau_{\rm f}$ to reach a fully fluidized state was reported to scale respectively as $\tau_{\rm f} \propto 1/\dot \gamma^\alpha$ or as $\tau_{\rm f} \propto 1/(\sigma-\sigma_\text{c})^\beta$, where $\alpha$ and $\beta$ are fluidization exponents that only depend on the material properties (see Fig.~\ref{fig1}). Interestingly, these two power laws naturally lead to a constitutive relation $\sigma$ {\it vs} $\dot \gamma$ given by the steady-state HB equation with an exponent $n=\alpha/\beta$ \cite{Divoux:2011b}. 

The above experimental findings have triggered a wealth of theoretical contributions aiming at reproducing long-lasting heterogeneous flows, some of which have successfully produced transient shear-banded flows together with non-trivial scaling laws for fluidization times \cite{Illa:2013,Moorcroft:2011,Moorcroft:2013,Hinkle:2016,Vasisht:2017,Liu:2018,Jain:2018}. While these contributions offer potential explanations for long-lasting transients, which appear to be age-dependent and related to structural heterogeneities \cite{Moorcroft:2011,Hinkle:2016,Vasisht:2017,Liu:2018,Liu:2018b}, none of these numerical studies captures the link between the exponents governing the transient regimes and that of the steady-state HB behavior.

From a more general perspective, shear banding has often been discussed as a first-order dynamical phase transition \cite{Dhont:1999,Lu:2000,Bocquet:2009,Chikkadi:2014,Divoux:2016}. In that framework, \textit{transient} shear banding can be interpreted as the coarsening of the fluid phase, which nucleates within the solid region and whose size $\delta$ can be seen as the growing length scale that characterizes the coarsening dynamics. In this letter, we show that the yielding transition and the corresponding transient shear-banding behavior can be described by a field theory based on a {``free energy''}, whose order parameter is the fluidity, i.e., the ratio between the shear rate and the shear stress. In such a theory, as first introduced by Bocquet \textit{et al.} \cite{Bocquet:2009} and later analyzed in Ref.~\cite{Benzi:2016}, shear-banded flows can be obtained as a minimum of a { ``free energy''} that depends on the fluidity and on the non-local, {i.e., spatially-dependent \cite{Dhont:1999,Lu:2000}},  rheological properties of the system.  {A link between the fluidity order parameter and the physics of elasto-plasticity at the mesoscale has been explored in Ref.~\cite{Nicolas:2013} based on Eshelby elastic response functions \cite{Eshelby:1957,Zaccone:2017,Dasgupta:2012}. Here we build upon the fluidity approach and extend it}, leading to analytical expressions for the scaling exponents $\alpha$ and $\beta$ that are in quantitative agreement with experiments and that provide a clear-cut explanation for the link between these exponents and the HB exponent $n$. Our findings demonstrate that non-local effects are key to understand transient shear banding in amorphous soft solids. 

\textit{Fluidity model.-} 
We start by considering that the bulk rheology of the system is governed by the dimensionless HB model,
$\Sigma=1+\dot{\Gamma}^{n}$, where $\Sigma=\sigma/\sigma_\text{c}$ is the shear stress normalized by the yield stress and  $\dot{\Gamma}=\dot\gamma/(\sigma_\text{c}/A)^{1/n}$ is the shear rate normalized by the characteristic frequency for the HB law. Given the spatial coordinate $y$ along the velocity gradient direction and the system size $L$, we next assume that the flow properties of the yield stress fluid are controlled by a {``free energy''} functional, $F[f] = \int_0^L \Phi(f,m,\xi)\, {\rm d}y$, where \cite{Bocquet:2009,Benzi_note1}
\begin{equation}\label{eq:bocquet}
\Phi(f,m,\xi) \equiv \left[ \frac{1}{2} \xi^2 (\nabla f)^2 - \frac{1}{2} m f^2 + \frac{2}{5} f^{5/2} \right]\,.
\end{equation}
The quantity $f=f(y)$ is the \textit{local} (dimensionless) fluidity defined by $f(y)=\dot{\Gamma}(y)/\Sigma(y)$ and represents the order parameter in the model. Following Refs.~\cite{Bocquet:2009,Benzi:2016}, $m^2$ is defined as:
\begin{equation}\label{eq:m}
m^2(\Sigma) \equiv\frac{(\Sigma-1)^{1/n}}{\Sigma}\, \hspace{.1in}\hbox{\rm for}\hspace{.1in} \Sigma\ge 1
\end{equation}
and $m^2=0$ for $\Sigma<1$. This formulation implies that, for $f(y)=m^2$ independently of $y$, the system flows homogeneously and follows the dimensionless HB model. Finally, the length scale $\xi$ is usually referred to as the ``cooperative'' scale and is of the order of a few times the size of the elementary microstructural constituents~\cite{Bocquet:2009,Goyon:2008,Goyon:2010,Geraud:2013,Geraud:2017}.
In steady-state, the flowing properties of the system can then be derived from the variational equation $\delta F/ \delta f = 0$. 
This equation predicts heterogeneous flow profiles as induced by wall effects but it cannot account for stable shear banding \cite{Benzi:2016}. Moreover, transient flow properties require that some temporal dynamics be specified for $f$. To overcome these limitations, we now generalize a recent theoretical proposal introduced in  Ref.~\cite{Benzi:2016} and apply it to describe transient flows.

\begin{figure*}[ht]
\centering
\includegraphics[width=0.7\linewidth]{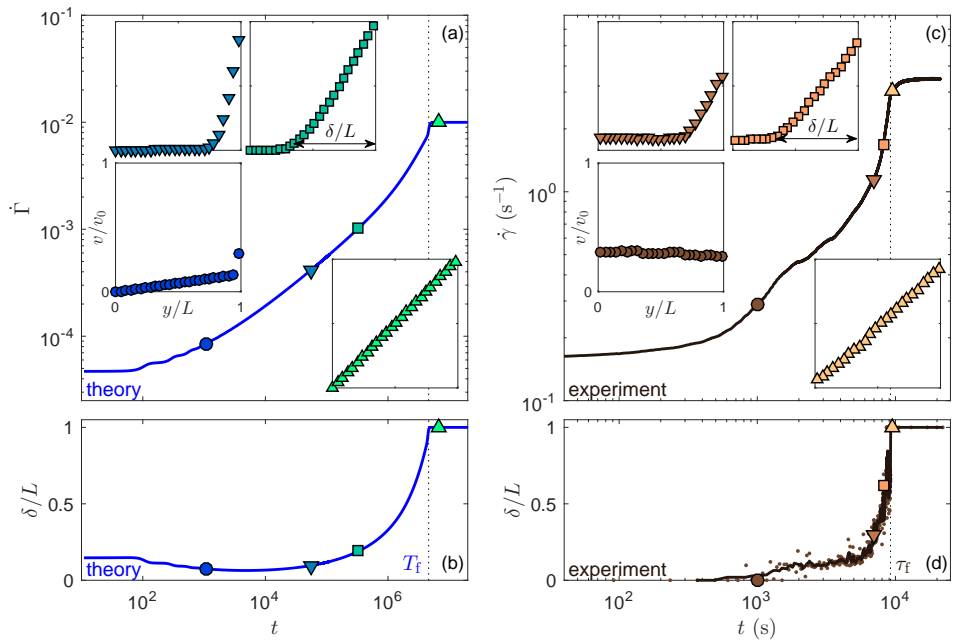}
\caption{(color online) Stress-induced fluidization dynamics in (a)-(b)~theory and (c)-(d)~experiment on a 1~\% wt.~carbopol microgel in a smooth concentric cylinder geometry of gap 1~mm. (a) and (c)~Shear rate $\dot\Gamma$ and $\dot\gamma$ vs time $t$ for a shear stress of $\Sigma=1.1$ and $\sigma=41$~Pa respectively. Insets: velocity profiles $v$ normalized by the velocity of the moving plate $v_0$ as a function of the distance $y$ to the fixed plate normalized by the gap size $L$. Profiles taken at different times [symbol, time]: (\textcolor{fig2a_1}{$\bullet$},1100); (\textcolor{fig2a_2}{$\blacktriangledown$},$5.5\,10^4$); (\textcolor{fig2a_3}{\tiny $\blacksquare$},$3.3\,10^5$); (\textcolor{fig2a_4}{$\blacktriangle$},$6.6\,10^6$) in (a) and (\textcolor{fig2b_1}{$\bullet$},1011~s); (\textcolor{fig2b_2}{$\blacktriangledown$},6927~s); (\textcolor{fig2b_3}{\tiny $\blacksquare$},8193~s); (\textcolor{fig2b_4}{$\blacktriangle$},9522~s) in (c). (b) and (d)~Width $\delta$ of the fluidized shear band normalized by the gap width $L$ vs time $t$. The vertical dashed lines crossing (a)-(b) and (c)-(d) respectively indicate the fluidization times $T_\text{f}$ and $\tau_\text{f}$.
\label{fig2}}
\end{figure*}

\textit{Stress-induced fluidization dynamics.-} Let us first focus on the yielding transition under an imposed shear stress $\sigma$ for which $m$ is a constant. We note that introducing $\tilde f = f / m^2 $ and $\tilde y = m^{1/2} y/\xi$ allows us to rescale homogeneously the functional $\Phi$ to $\Phi(f,m,\xi) =  m^5 \tilde \Phi (\tilde f) $, where~\cite{Benzi_note2}
\begin{equation}\label{eq:bocquet_normalized}
 \tilde \Phi (\tilde f) =\left[ \frac{1}{2} (\tilde\nabla \tilde{f})^2 - \frac{1}{2} \tilde{f}^2 + \frac{2}{5} \tilde{f}^{5/2} \right]\,.
\end{equation}
The advantage of using $\tilde f $ and $\tilde y$  is that we can now formulate the dynamical equation independently of both the strength of external forcing $m$ and $\xi$. We further assume that the system reaches a stable equilibrium configuration corresponding to a minimum of $F[\tilde f]$ and that such dynamics is governed by a ``mobility'' $k(\tilde{f})$, for which the most general dynamical equation reads \cite{Benzi_note1}
\begin{equation}\label{new1}
\begin{split}
\frac{\partial \tilde f}{\partial t} & = - m^5 k(\tilde{f}) \frac{\delta F[\tilde f]}{\delta \tilde f}  \\
& = m^5 k( \tilde f) \left[ \tilde  \Delta \tilde f + \tilde f - \tilde f ^ {3/2}\right]\,.
\end{split}
\end{equation}
If the mobility $k(\tilde f)$ is an analytic function of $\tilde f$ and $k(0)= 0$, then Eq.~(\ref{new1}) can account for a shear-banding solution in the general form $ \tilde f(\tilde y) = 0$ (solid branch) for $ \tilde y \in [0, \tilde L-\tilde \delta]$ and $\tilde f(\tilde y)$ solution of $\tilde \Delta \tilde f + \tilde f - \tilde f ^ {3/2} = 0$ (fluidized branch) for $\tilde y  \in [\tilde L-\tilde\delta,\tilde L]$, where $\tilde\delta$ is the rescaled size of the fluidized region. Furthermore, transient shear banding corresponds to the case where the solid branch $\tilde f=0$ is an unstable solution. 
To explore this latter case, we next consider the time dynamics in  Eq.~\eqref{new1} with $k(\tilde f) = \tilde f$ and fixed initial conditions. Note that the initial conditions influence mainly the early-time response of the fluid. {A detailed discussion on the choice of $k(\tilde f)$ and on intial conditions is given in the Supplemental Material.}
Equation~\eqref{new1} is solved numerically for $\Sigma=1.1$ and $\xi/L=0.01$ in Figs.~\ref{fig2}(a)-(b), assuming $\tilde f (\tilde y,0) = \tilde f_0 \ll 1$ for the initial solid-like state and $\tilde f(\tilde{L},t) = 1$ and $\partial_{ \tilde y} \tilde f (0, t)=0$ for boundary conditions at the two different walls. Such a choice will be addressed below in the discussion section. As seen in the velocity profiles $v(y)$ [insets in Fig.~\ref{fig2}(a)], the system forms a shear band near $y=L$ at time $ t>0$. The shear band grows in time and the system eventually reaches the stable equilibrium configuration $\tilde f(\tilde y,t)=1$ within a well-defined fluidization time $T_\text{f}$. This phenomenology is in remarkable agreement with experimental observations in Figs.~\ref{fig2}(c) and (d) for a carbopol microgel. In particular, the band size $\delta(t)$ follows very similar growths whatever the applied stress (see Supplemental Fig.~S1). 

\begin{figure}[t!]
\centering
\includegraphics[width=0.8\columnwidth]{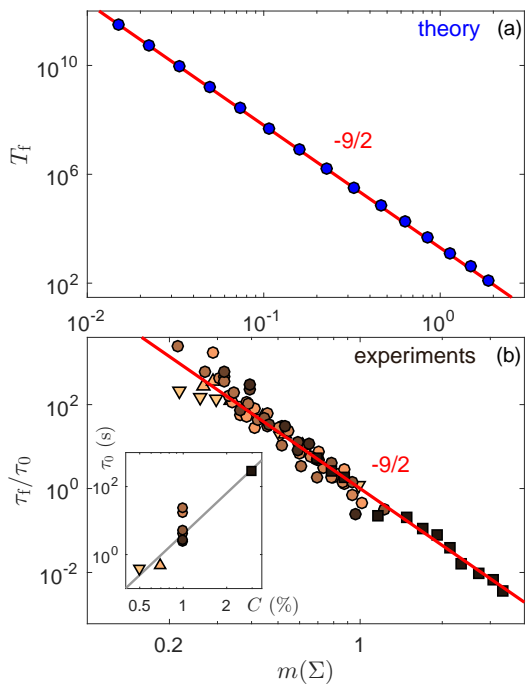}
\caption{(color online) Stress-induced fluidization time as a function of $m(\Sigma)$ defined by Eq.~\eqref{eq:m}. (a)~Theoretical predictions $T_\text{f}$. (b)~Experiments from Fig.~\ref{fig1} where each data set for $\tau_\text{f}$ was rescaled by the time $\tau_0$ shown in the inset as a function of the microgel concentration $C$ (see also Supplemental Table~S1). Red lines show the predicted power law with exponent $-9/2$. The best power-law fits of the whole data sets yield exponents $-4.46\pm0.10$ and $-4.69\pm 0.33$ respectively for theory and experiments. The gray line in the inset is $\tau_0\sim C^4$.
\label{fig3}}
\end{figure}

Using Eq.~(\ref{new1}), we may predict the scaling behavior of the fluidization time $T_\text{f}$ as a function of $m$. Upon rescaling the time as $\tilde t = m^5 t$, we observe that Eq.~(\ref{new1}) no longer depends on $m$.  Regardless of the specific function $k(\tilde f)$, we expect that the shear band expands with some characteristic velocity $\tilde v_\text{f}$ independent of $m$. Therefore, the rescaled fluidization time should be proportional to $ \tilde L / \tilde v_\text{f}$. It follows that the fluidization time should exhibit the scaling  $T_\text{f} \sim \tilde L / (m^5 \tilde v_\text{f}) \sim 1/ (\xi m^{9/2})$ {independently of the specific functional form of $k(\tilde f)$}. The numerical integration of Eq.~(\ref{new1}) for various values of $m$ leads to the fluidization times $T_\text{f}$ shown in Fig.~\ref{fig3}(a), which nicely follow the predicted $m^{-9/2}$ power-law decay. Such a scaling is also in excellent agreement with the experimental data of Fig.~\ref{fig1} when rescaled and plotted in terms of $m(\Sigma)$ based on the experimental steady state HB parameters [see Fig.~\ref{fig3}(b) and discussion below].

\begin{figure}[t!]
\centering
\includegraphics[width=0.8\linewidth]{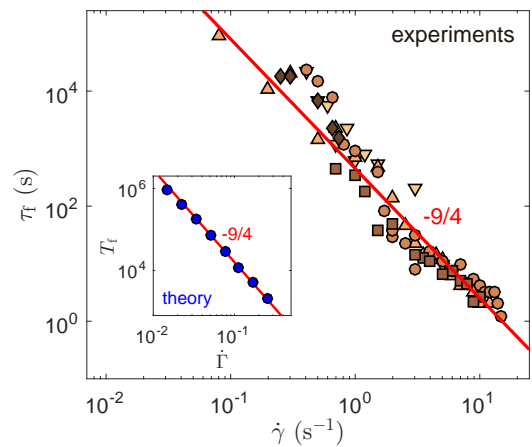}
\caption{(color online) Strain-induced fluidization time $\tau_\text{f}$ vs shear rate $\dot\gamma$ for a 1\%~wt. carbopol microgel under the various experimental conditions listed in Supplemental Table~S2. Inset: theoretical prediction for $T_\text{f}$ vs $\dot\Gamma$. Red lines show the predicted power law with exponent $-9/4$. The best power-law fits of the whole data sets yield exponents $-2.15\pm 0.10$ and $-2.45\pm0.23$ respectively for theory and experiments. 
\label{fig4}}
\end{figure}

\textit{Strain-induced fluidization.-} We now proceed to show that the same approach allows us to rationalize the yielding transition under an imposed shear rate $\dot\Gamma$. In that case, we must supplement the theory by the fluidity equation $\dot \Sigma = \dot \Gamma - f \Sigma$, which corresponds to a single Maxwell mode for the evolution of the stress \cite{Moorcroft:2011}. Moreover, $m$ being a function of time, we can no longer use the rescaling $\tilde f = f / m^2$. Since $\dot \Gamma$ is a constant, we rather introduce the rescaled variable $\tilde f = f /\dot \Gamma$. Upon rescaling the spatial variable as $\tilde y = \dot \Gamma^{1/4} y/ \xi$, the analogous of Eq.~(\ref{new1}) reads
\begin{equation}
\label{15}
\frac{\partial \tilde f}{\partial t}  = \dot \Gamma^{5/2 }k(\tilde f) \left[  \tilde \Delta \tilde f + \tilde m \tilde f - \tilde f^{3/2} \right]\,,
\end{equation}
where $\tilde m = m/\dot \Gamma^{1/2}$. Under the assumption that $\tilde m$ remains roughly constant during the shear band evolution, rescaling time as $\tilde t =\dot\Gamma^{5/2} t$ leads to $T_\text{f}  \sim \tilde L /( \dot \Gamma^{5/2}\tilde v_\text{f})   \sim   1/(\xi\dot \Gamma^{9/4})$. 
The inset of Fig.~\ref{fig4} shows the actual $T_\text{f}$ computed numerically from Eq.~(\ref{15}) with $k(\tilde f) = \tilde f$ for different shear rates $\dot \Gamma$. The results are very well fitted by a power-law decay of exponent $2.15\pm 0.10$, quite close to the theoretical exponent $\alpha=9/4$, and in good agreement with experiments on a 1\%~wt. carbopol microgel for various geometries and boundary conditions that lead to an exponent of $2.45\pm 0.23$ (see Fig.~\ref{fig4} and Supplemental Table~S2).

\textit{Discussion.-} Let us now compare the theoretical findings against experimental data. Coming back to the case of an imposed shear stress and to the definition of $m$ in Eq.~\eqref{eq:m}, we note that $T_\text{f}\sim m^{-9/2}$ corresponds to the scaling $T_\text{f}\sim (\Sigma-1)^{-9/4n}$ in terms of the reduced viscous stress $\Sigma-1$. This corresponds to a fluidization exponent $\beta=9/4n$. To illustrate such a scaling, numerical results are plotted in Supplemental Fig.~S2 for different values of $n$ covering the range reported in experiments ($n\simeq 0.30$--0.57). The spread of the exponents $\beta\simeq 3$--8 nicely corresponds to that observed experimentally ($\beta\simeq 2.8$--6.2). More specifically, these theoretical predictions prompt us to revisit the experimental data shown in Fig.~\ref{fig1} by computing estimates of $m(\Sigma)$ using Eq.~\eqref{eq:m} with $\Sigma=\sigma/\sigma_\text{c}$ and the HB parameters $\sigma_\text{c}$ and $n$ determined at steady state \cite{Divoux:2011b}. When plotted as a function of $m(\Sigma)$, the experimental fluidization times remarkably collapse onto the predicted scaling $\tau_\text{f}\sim m(\Sigma)^{-9/4}$, provided $\tau_\text{f}$ is rescaled by a characteristic time $\tau_0$ independent of the applied stress [see Fig.~\ref{fig3}(b)]. Although a clear physical interpretation of $\tau_0$ is still lacking  \cite{Benzi_note3}, the collapse of the experimental data seen in Fig.~\ref{fig3}(b) is a strong signature of the predictive power of the theory.

Another key outcome of the proposed approach is that, assuming an underlying HB rheology, it provides the first theoretical analytical expressions for both fluidization exponents $\alpha$ and $\beta$, in quantitative agreement with experimental results. Moreover, the ratio of these exponents, $\alpha / \beta = (9/4)/(9/4n)= n$, coincides with the Herschel-Bulkley exponent exactly as in experiments \cite{Divoux:2011b,Divoux:2012}. Therefore, the present theory provides a natural framework for justifying the empirical connection between transient and steady-state flow behaviors. 

Furthermore, the scaling found here for $\tau_\text{f}$ is extremely robust and depends only weakly on the initial conditions. As illustrated in Supplemental Figs.~S3 and S4 for two different initial values of the fluidity in the gap, the shear rate either shows a monotonic increase up to complete fluidization or displays a decreasing trend with a well-defined minimum before increasing towards steady state. Yet, the fluidization time remains comparable in both cases. Note also that, at early stage, $\dot \Gamma$ shows a power-law decrease in time that is strongly reminiscent of the primary creep regime reported in amorphous soft materials \cite{Bauer:2006,Divoux:2011b,Grenard:2014,Leocmach:2014,Helal:2016,Lidon:2017,Aime:2018}. In the present model, the power-law exponent may take any value between $-2/3$ and $0$ depending on the choice of $k(\tilde f)$, thus providing an explanation for the diversity of exponents reported in the literature.

To conclude, our results show that the {``free energy''} approach originally introduced to account for non-local effects in steady-state flows of complex fluids \cite{Bocquet:2009} also captures long-lasting transient heterogeneous flows: thanks to cooperative effects, a fluidized band nucleates and grows until complete yielding, which quantitatively matches the experimental phenomenology. In this framework, transient shear banding appears as the dynamical signature of the unstable nature of the solid branch at $\dot\gamma=0$ in the flow curve \cite{Varnik:2003,Varnik:2004,Bonn:2017}. More generally, as explored in Ref.~\cite{Benzi:2016}, the present model also accounts for steady-state shear banding when cooperative effects are hindered, e.g.,~by mechanical noise that prevents the shear band from growing through cascading plastic events. Such a connection between transient and steady-state behaviors in terms of cooperativity-induced stability of the shear band offers for the first time a unified framework for describing the local scenario associated with the yielding dynamics of soft glassy materials.  

\begin{acknowledgments}
The authors thank David Tamarii for help with the experiments as well as Emanuela Del Gado and Suzanne Fielding for fruitful discussions. This research was supported in part by the National Science Foundation under Grant No. NSF PHY~17-48958 through the KITP program on the Physics of Dense Suspensions.
\end{acknowledgments}


\begin{thebibliography}{53}
\bibitem [{Ben({\natexlab{a}})}]{Benzi_note1}%
  \BibitemOpen
  \href@noop {} {} {\natexlab{a}}\ \bibinfo {note} {{F}or simplicity, we
  shall use the same symbols $y$ and $t$ for the spatial and temporal degrees
  of freedom both in experiments and theory.}\BibitemShut {Stop}%
\bibitem [{Ben({\natexlab{b}})}]{Benzi_note2}%
  \BibitemOpen
  \href@noop {} {} {\natexlab{b}}\ \bibinfo {note} {{T}he tildes over the
  {gradient} $\nabla$ and {Laplacian} $\Delta$ operators in
  Eqs.~\eqref{eq:bocquet_normalized}, \eqref{new1} and \eqref{15} indicate that
  derivatives are taken over the rescaled spatial variable $\tilde
  y$.}\BibitemShut {Stop}%
\bibitem [{Ben({\natexlab{c}})}]{Benzi_note3}%
  \BibitemOpen
  \href@noop {} {} {\natexlab{c}}\ \bibinfo {note} {{T}he rescaling factor
  $\tau_0$ strongly depends on the system concentration $C$, scaling roughly as
  $C^4$ as shown in the inset of Fig.~\ref{fig3}(b), and does not appear to be
  trivially linked to the characteristic time $(A/\sigma_\text{c})^{1/n}$ that
  one can extract from the HB behavior, which only varies from 0.12~s to 0.61~s
  in our experiments.}\BibitemShut {Stop}%
\end{thebibliography}

%


\clearpage
\newpage
\onecolumngrid
\setcounter{page}{1}

\begin{center}
    {\large\bf Unified theoretical and experimental view on transient shear banding.}
\end{center}

\begin{center}
    {\large\bf Supplementary information}
\end{center}

\section{Experimental parameters}

\setcounter{equation}{5}

\setcounter{figure}{0}
\global\def\thefigure{S\arabic{figure}}
\setcounter{table}{0}
\global\def\thetable{S\arabic{table}}

\begin{table}[htb]
\begin{tabular}{c|c|c|c|c|c|c|c|c|c}
Symbol & $C$ ($\%$) & Geometry & BC & $L$ (mm) & $\sigma_\text{c}$ (Pa) & $n$ & $A$ (Pa.s$^n$) & $\beta$ & $\tau_0$ (s)\\
\hline\hline
\textcolor{fig1_1}{$\blacktriangledown$} & 0.5 & parallel plate & rough & 1 & 21.8 & 0.57 & 9.1 & 2.8 & 2.5\\   
\textcolor{fig1_2}{$\blacktriangle$} & 0.7 & parallel plate & rough & 1 & 32.9 & 0.54 & 12.3 & 3.3 & 2.0 \\   
\textcolor{fig1_3}{$\bullet$} & 1 & cone \& plate & smooth & - & 30.0 & 0.50 & 10.6 & 4.2 & 0.25 \\   
\textcolor{fig1_4}{$\bullet$} & 1 & concentric cylinders & rough & 1.1 & 27.8 & 0.53 & 11.3 & 4.2 & 0.06 \\   
\textcolor{fig1_5}{$\bullet$} & 1 & concentric cylinders & smooth & 1 & 30.4 & 0.53 & 10.3 & 4.9 & 0.04 \\   
\textcolor{fig1_6}{$\bullet$} & 1 & parallel plate & smooth & 1 & 40.2 & 0.43 & 20.8 & 4.5 & 0.2 \\   
\textcolor{fig1_7}{$\bullet$} & 1 & parallel plate & rough & 1 & 47.4 & 0.50 & 18.7 & 4.5 & 0.4 \\   
\textcolor{fig1_8}{$\bullet$} & 1 & parallel plate & rough & 3 & 47.4 & 0.50 & 18.7 & 5.9 & 0.35 \\   
\textcolor{fig1_9}{\tiny $\blacksquare$} & 3 & parallel plate & rough & 1 & 115.5 & 0.30 & 99.7 & 6.2 & 3.3\,$10^{-3}$ \\   
\end{tabular}
\caption{Experimental parameters for stress-induced fluidization of carbopol microgels of weight concentration $C$ in different shearing geometries with different boundary conditions (BC) and gap widths $L$. The yield stress $\sigma_\text{c}$, the shear-thinning exponent $n$ and the consistency index $A$ are inferred from Herschel-Bulkley fits of the steady-state $\sigma$ vs $\dot\gamma$ data. $\beta$ is the exponent of the best power-law fit of the stress-induced fluidization time $\tau_\text{f}$ vs $\sigma-\sigma_\text{c}$ shown in Fig.~1. $\tau_0$ is the characteristic time used to rescale $\tau_\text{f}$ in Fig.~3(c). For a fixed weight concentration of 1~\%, it varies by one order of magnitude depending on the batch sample, on the geometry and on boundary conditions. This suggests a subtle dependence of $\tau_0$ on the microscopic details of the system and its interaction with the shearing walls, standing out as an open issue. The symbols in the first column are those used in Fig.~1 and Fig.~3(c) in the main text.} \label{table1}
\end{table}

\begin{table}[htb]
\begin{tabular}{c|c|c|c|c|c|c|c|c|c}
Symbol & Geometry & BC & $L$ (mm) & $\alpha$ \\
\hline\hline 
\textcolor{fig4_1}{$\blacktriangledown$} & concentric cylinders & smooth & 0.5 & 2.6\\
\textcolor{fig4_2}{$\blacktriangle$} & concentric cylinders & rough & 1.1 & 2.3\\   
\textcolor{fig4_3}{$\bullet$} & concentric cylinders & smooth & 1.5 & 2.5\\   
\textcolor{fig4_4}{\tiny $\blacksquare$} & concentric cylinders & smooth & 3 & 2.0 \\   
\textcolor{fig4_5}{$\blacklozenge$} & cone \& plate & smooth & - & 2.3 \\
\end{tabular}
\caption{Experimental parameters for strain-induced fluidization of a 1\%~wt. carbopol microgel in different shearing geometries with different boundary conditions (BC) and gap widths $L$. $\alpha$ is the exponent of the best power-law fit of the strain-induced fluidization time $\tau_\text{f}$ vs $\dot\gamma$ found for each individual data set. The symbols in the first column are those used in Fig.~4 in the main text.} \label{table2}
\end{table}

The experimental conditions leading to the results shown in Fig.~1, Fig.~2(c) and (d), Fig.~3(b) and Fig.~4 in the main text are gathered in  Tables~\ref{table1} and \ref{table2}. In all cases, carbopol microgels were prepared at a weight concentration $C$ following the protocol described in Ref.~\cite{Divoux:2011b}. As explored in Refs.~\cite{Baudonnet:2002,Baudonnet:2004,Lee:2011,Geraud:2013,Geraud:2017}, the details of the preparation protocol, especially the carbopol type, the final pH and the mixing procedure, have a strong impact on the microstructure of the resulting microgels and on their rheological properties. In particular, carbopol microgels prepared with a similar procedure as the present samples \cite{Geraud:2013,Geraud:2017} were shown to be constituted of jammed, polydisperse swollen polymer particles of typical size $6~\mu$m. The cooperative length $\xi$ was estimated to be about 2 to 5 times the particle size thanks to local rheological measurements in microchannels \cite{Geraud:2013,Geraud:2017}.

The samples are loaded in a shearing cell attached to a standard rheometer (Anton Paar MCR301). Experiments listed in Tables~\ref{table1} and \ref{table2} performed in parallel-plate and in concentric-cylinder geometries with gaps larger than 0.5~mm have already been described at length in Refs.~\cite{Divoux:2010,Divoux:2011b,Divoux:2012}. The present work also includes new data sets obtained in a smooth cone-and-plate geometry (steel cone of diameter 50~mm, angle 2$^\circ$, truncation 55~$\mu$m) and in a smooth concentric-cylinder geometry of gap 0.5~mm (Plexiglas cylinders, outer diameter 50~mm, height 30~mm). Note that the HB parameters $\sigma_\text{c}$, $A$ and $n$ for measurements in parallel-plate geometries were extracted from the steady-state rheological data, which explains the differences in the yield stress (and thus in the exponent $\beta$) indicated in Table~\ref{table1} and in Ref.~\cite{Divoux:2011b} where $\sigma_\text{c}$ was directly extracted from the $\tau_\text{f}$ vs $\sigma$ data.

Under an imposed shear stress, the fluidization time $\tau_\text{f}$ was shown to correspond to the last inflection point of the shear rate response $\dot\gamma(t)$ \cite{Divoux:2011b}. This allows us to measure $\tau_\text{f}(\sigma)$ in the absence of simultaneous velocity measurements, e.g., in cone-and-plate and in parallel plate geometries. As for experiments performed under an imposed shear rate, the end of the transient shear-banding regime is associated with a significant drop in the stress response $\sigma(t)$ \cite{Divoux:2010,Divoux:2012} that is used to estimate $\tau_\text{f}(\dot\gamma)$ in the cone-and-plate geometry. 

\begin{figure*}[htb]
\centering
\includegraphics[width=0.8\linewidth]{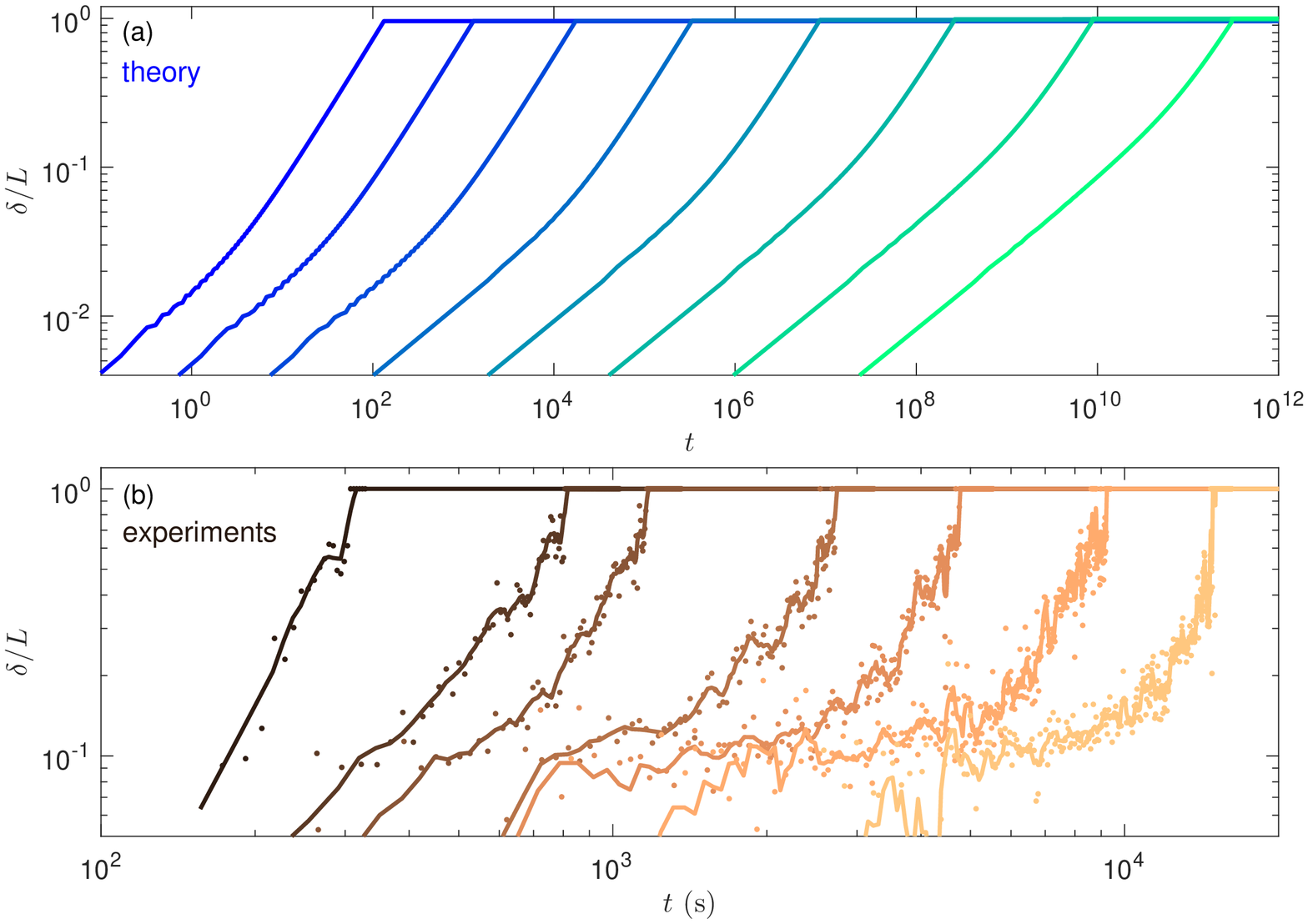}
\caption{Width $\delta$ of the stress-induced fluidized shear band normalized by the gap width $L$ vs time $t$ in (a)~theory for $\Sigma=1.015$, 1.034, 1.076, 1.171, 1.384, 1.865, 2.946 and 5.379 from right to left and (b)~experiments for $\sigma=39$, 41, 42, 44, 45.5, 47 and 50~Pa from right to left. Experiments performed on a 1\%~wt. carbopol microgel in a smooth concentric cylinder geometry with gap width $L=1$~mm. The solid lines show a smoothed version of the raw data (colored {\tiny $\bullet$}) using a moving average over 5 successive data points.
\label{suppfig1}}
\end{figure*}

In the case of concentric cylinders, rheological measurements are supplemented by time-resolved local velocity measurements. The technique is based on the scattering of ultrasound by hollow glass microspheres (Potters, Sphericel, mean diameter 6~$\mu$m, density 1.1) suspended at a volume fraction of 0.5~\% within the carbopol microgel. It was previously shown that such seeding of the microgel samples does not affect their fluidization dynamics \cite{Divoux:2011b}. Full details on ultrasonic velocimetry coupled to rheometry can be found in Ref.~\cite{Manneville:2004a}. This technique outputs the tangential velocity $v(y,t)$ as a function of the distance $y$ to the fixed wall and as a function of time $t$. The outer fixed cylinder is thus located at $y=0$ and the inner rotating cylinder at $y=L$, where $L$ is the width of the gap between the two cylinders. Fig.~2(c) in the main text shows a few velocity profiles $v(y,t)/v_0(t)$ vs $y/L$ where the velocity is normalized by the current velocity $v_0(t)$ of the moving wall deduced from the shear rate response $\dot\gamma(t)$. Each velocity profile is itself an average over 10 to 1000 successive velocity measurements, which corresponds typically to an average over 8~s to 140~s. The typical standard deviation of these measurements is about the symbol size. Note that these data, obtained in a smooth geometry, show significant wall slip, as opposed to those shown in Ref.~\cite{Divoux:2011b} for rough boundary conditions. Finally, each individual velocity profile is fitted by linear functions over $y$-intervals extending respectively within the solid-like region and within the fluidized band (when present). The intersection of the two fits yields the width $\delta$ of the fluidized band as shown in Fig.~2(c) and as plotted as a function of time in Figs.~2(d) and \ref{suppfig1}(b).

\section{Theoretical considerations}

{In this section, we examine in more details some theoretical aspects concerning the fluidity model used in the main text in order to justify our choice of function $k(\tilde{f})$. We specifically address the basic differences between the general case $k(\tilde f) = \tilde f^p$ with $p>0$ [hereafter referred to as case~(I)] and the particular case $k(\tilde f) = \textrm{const}$ [hereafter referred to as case~(II)]. As already outlined in Ref.~\cite{Benzi:2016}, case (I) admits stationary solutions with the coexistence of two rheological branches: the solid branch where $\tilde f = \tilde f_s = 0$ and a fluid branch $\tilde f = \tilde f_b > 0$. In other words, case (I) admits for stationary solution a shear-banded profile whilst this cannot be for case (II). Such a difference matters because these two fluidization mechanisms yield different time scales. Indeed,} assuming that the initial condition $\tilde f(0)$ is homogeneous and neglecting the term $\tilde \Delta \tilde f$  in Eq.~(4), we obtain

\begin{equation}
\label{new22}
\frac{\partial \tilde f}{\partial t} = m^5 k(\tilde f)  \left[ \ \tilde f - \tilde f ^ {3/2} \right ]  \,.
\end{equation}
{We further consider the short time behavior of the instability by neglecting the term $\tilde f^{3/2}$ in Eq.~(\ref{new22}). It is enough to compare the two cases for the choice $p = 1$. For case (I), we obtain:
\begin{equation} 
\label{new81}
 \tilde f(t) = \frac{\tilde f(0)}{1-m^5\tilde f(0) t}  \,,
 \end{equation}
 while for case (II) we get
\begin{equation}
\label{new82}
\tilde f(t) = \tilde f(0)\exp(m^5 t) \,.
\end{equation}
Upon comparing Eqs.~(\ref{new81}) and (\ref{new82}), it is clear that the characteristic time for the instability depends on the initial condition $\tilde f(0)$ for case (I), while it is independent of the initial condition for case (II).
This dependence on $\tilde f(0)$ for case (I) probably explains the small yet detectable dependence of the fluidization time $T_f$ on the initial condition as reported in Fig.~\ref{figvel}. There, assuming two different initial conditions, we show that 
\begin{equation}
\label{new83}
\frac{T_{\text{f},1}}{T_{\text{f},2}} = C_1-C_2 \log \left[m(\Sigma)\right]\,,
\end{equation}
where $T_{\text{f},i}$ is the fluidization time computed for initial condition $i$ and $C_1$ and $C_2$ are positive constants. This is not observed for case (II), whose fluidization time is independent on the initial condition since Eq.~(4) for case (II) is essentially a reaction-diffusion equation \cite{Crank:1979,Murray:2003}.}

Finally, we discuss how cases (I) and (II) differ in the decay rate of the fluidity. {Indeed, for a sufficiently large initial fluidity, the term $\tilde f ^ {3/2}$ is dominant in Eq.~(\ref{new22}) so that the fluidity decreases.} 
The relaxation equation thus takes the following form
\begin{equation}\label{new24}
\frac{\partial \tilde f}{\partial t} {= -m^5} k( \tilde f)  \tilde f ^ {3/2} = -{m^5}\tilde f ^ {p+3/2}\,,
\end{equation}
{with $p>0$ for case (I) and $p=0$ for case (II).} The solution of Eq.~(\ref{new24}) reads
\begin{equation} \label{new26}
\tilde f(t) = \frac{A}{ (1 + B t)^b}\,,
\end{equation}
where $b = 2/(1+2p)$ and $A$ and $B$ are suitable constants. For $p=1$, one has $b = 2/3$ as already discussed in the main text. {This corresponds to the scaling observed experimentally for the shear rate (or fluidity) response under a constant stress in Ref.~\cite{Divoux:2011b}, which motivates our choice of $p=1$}. Note that for case (II) we obtain an exponent $b = 2$ far away from any experimental finding {\cite{Bauer:2006,Siebenburger:2012a,Leocmach:2014,Grenard:2014,Helal:2016,Lidon:2017,Aime:2018}}.

{The above discussion around Eq.~(\ref{new22}) leads to two interesting conclusions. First, the growth of the instability depends on the initial conditions for case (I) but not for case (II). The weak dependence of fluidization times on initial conditions for case (I) could also be linked to the logarithmic dependence of $T_\text{f}$ on the waiting time spent at rest as reported in Ref.~\cite{Benzi:2016} although a thorough comparison of aging effects in theory and experiments is left for future work. Second, the decay of the fluidity is an indication of the functional form of the mobility function $k$ and points to a linear behavior of $k(\tilde f)$.}

{In summary, complex materials as the one considered in this Letter show a broad spectrum of relaxation time scales, which cannot be reduced to a simple diffusion constant. This simple argument allows us to rule out case (II) where $k(\tilde f) = \textrm{const}$ would correspond to a single relaxation time. Indeed, although case (II) predicts the same scaling behavior for the fluidization time as case (I), it fails to reproduce several key features of the experimental results on carbopol microgels. This is the reason why we chose to use $k(\tilde f)= \tilde f^p$ with $p=1$ in the main text.}

\newpage
\section{Supplemental figures}

\begin{figure}[t!]
\centering
\includegraphics[width=0.4\columnwidth]{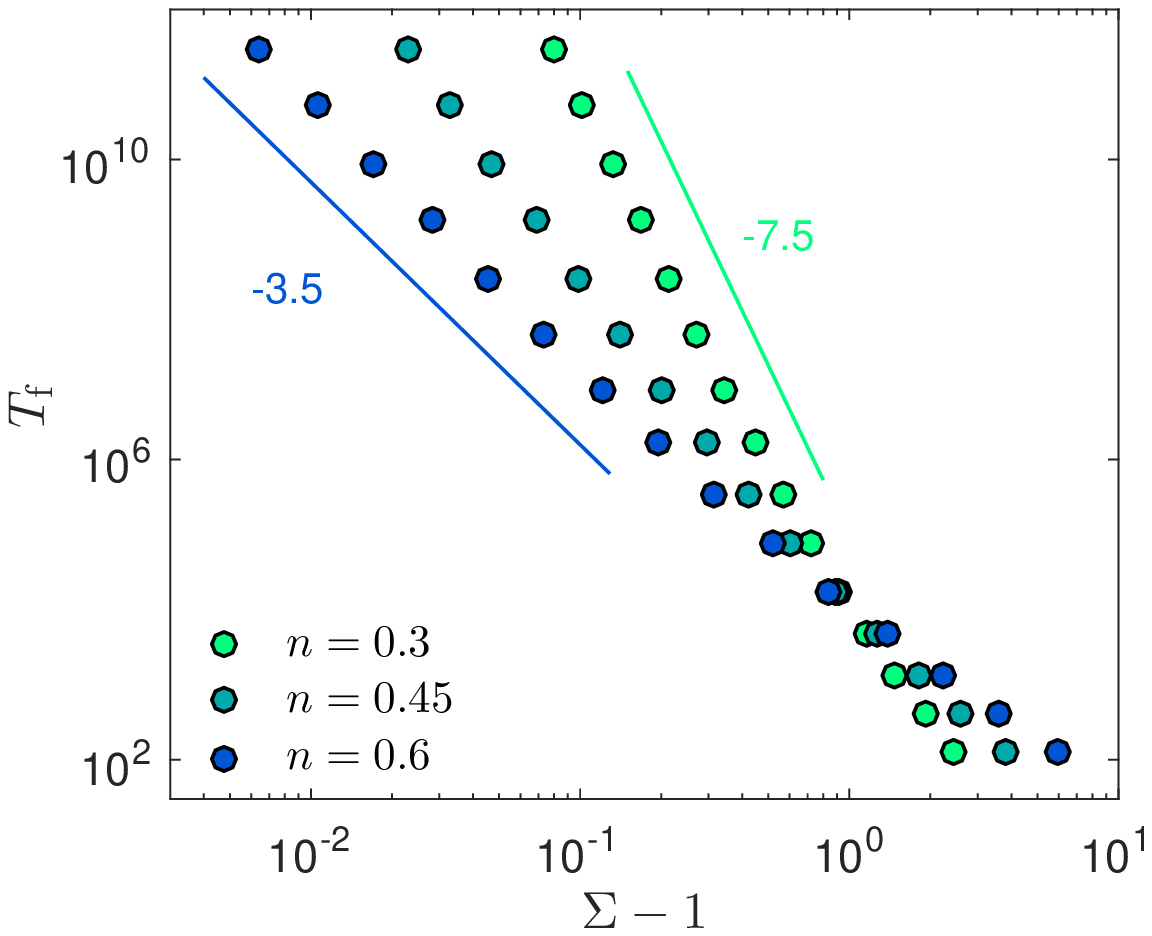}
\caption{Theoretical predictions for the stress-induced fluidization time $T_\text{f}$ as function of the reduced stress $\Sigma-1$ for three different values of the Herschel-Bulkley exponent ($n=0.3, 0.45$ and $0.6$). Solid lines show power laws with exponents -3.5 and -7.5.
\label{suppfig2}}
\end{figure} 

\begin{figure*}[htb]
\centering
\includegraphics[width=0.8\linewidth]{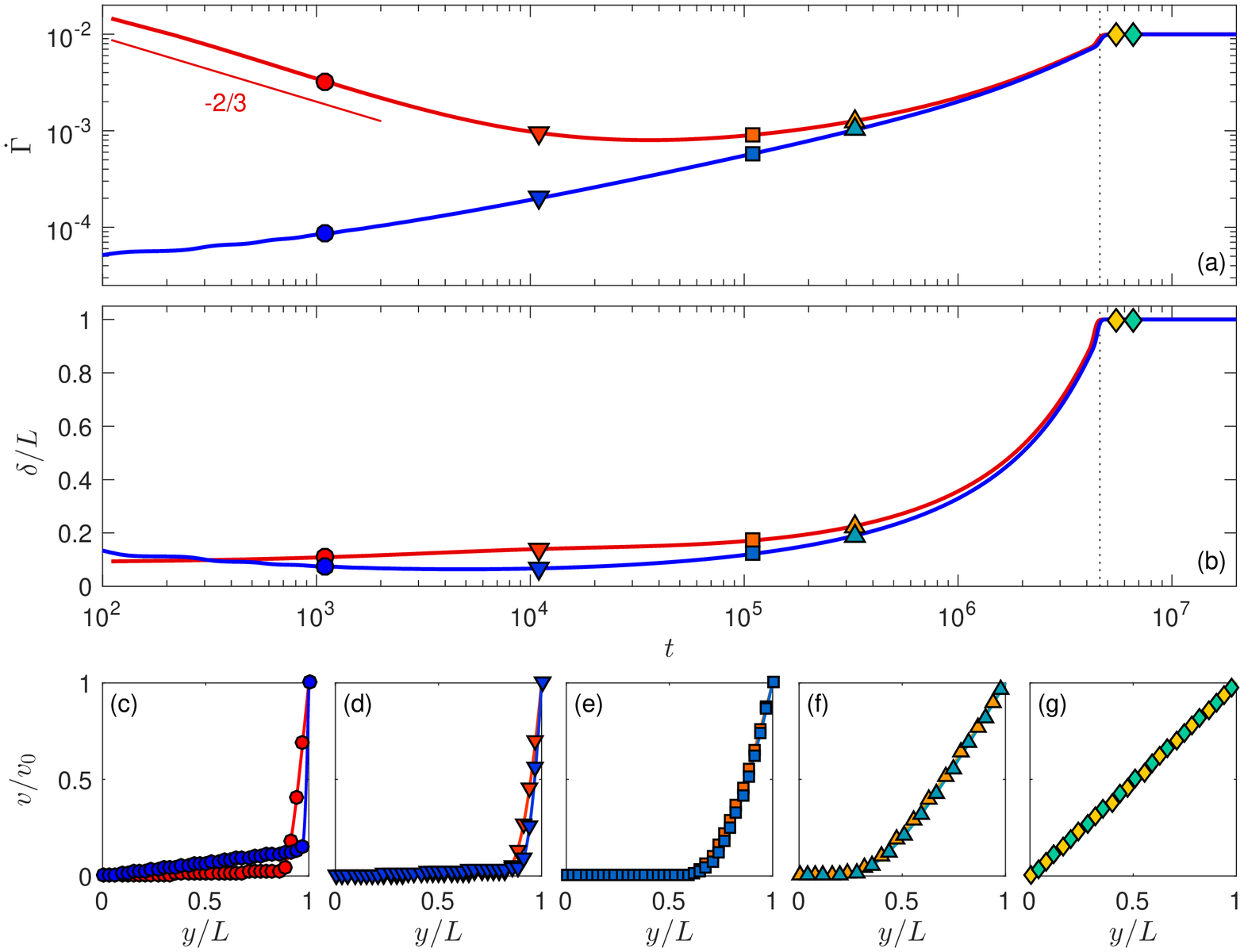}
\caption{Theoretical predictions for stress-induced fluidization dynamics for $\Sigma=1.1$ under two different initial conditions {with $k(\tilde{f})=\tilde{f}$}. (a)~Shear rate $\dot \Gamma$ and (b)~width $\delta$ of the fluidized shear band vs time $t$. The blue line corresponds to the data shown in Fig.~2(a) obtained with the initial condition $\tilde f(\tilde y,0)=\tilde{f}_0=2.5\,10^{-5}$. The red line is obtained with an initial condition where part of the material is solid-like, $\tilde f(\tilde y,0)=\tilde{f}_0=2.5\,10^{-5}$ for $0<\tilde{y}<0.9\,\tilde{L}$, while the rest of the material for $0.9\leq\tilde{y}\leq\,\tilde{L}$ is already fluidized with a fluidity that is 10 times the one predicted by the HB law. The vertical dashed lines indicate the fluidization time. (c)--(g)~Normalized velocity profiles $v(r)$ taken at different times [symbols, time]: (\textcolor{sfig2a_1}{$\bullet$},\textcolor{sfig2b_1}{$\bullet$},1100); (\textcolor{sfig2a_2}{$\blacktriangledown$},\textcolor{sfig2b_2}{$\blacktriangledown$},$1.1\,10^4$); (\textcolor{sfig2a_3}{\tiny $\blacksquare$},\textcolor{sfig2b_3}{\tiny $\blacksquare$},$1.1\,10^5$); (\textcolor{sfig2a_4}{$\blacktriangle$},\textcolor{sfig2b_4}{$\blacktriangle$},$3.3\,10^5$); (\textcolor{sfig2a_5}{$\blacklozenge$},$5.5\,10^6$) and (\textcolor{sfig2b_5}{$\blacklozenge$},$6.6\,10^6$).
\label{suppfig3}}
\end{figure*} 

\clearpage

\begin{figure}[h]
	\begin{center}
		\includegraphics[width=0.4\columnwidth]{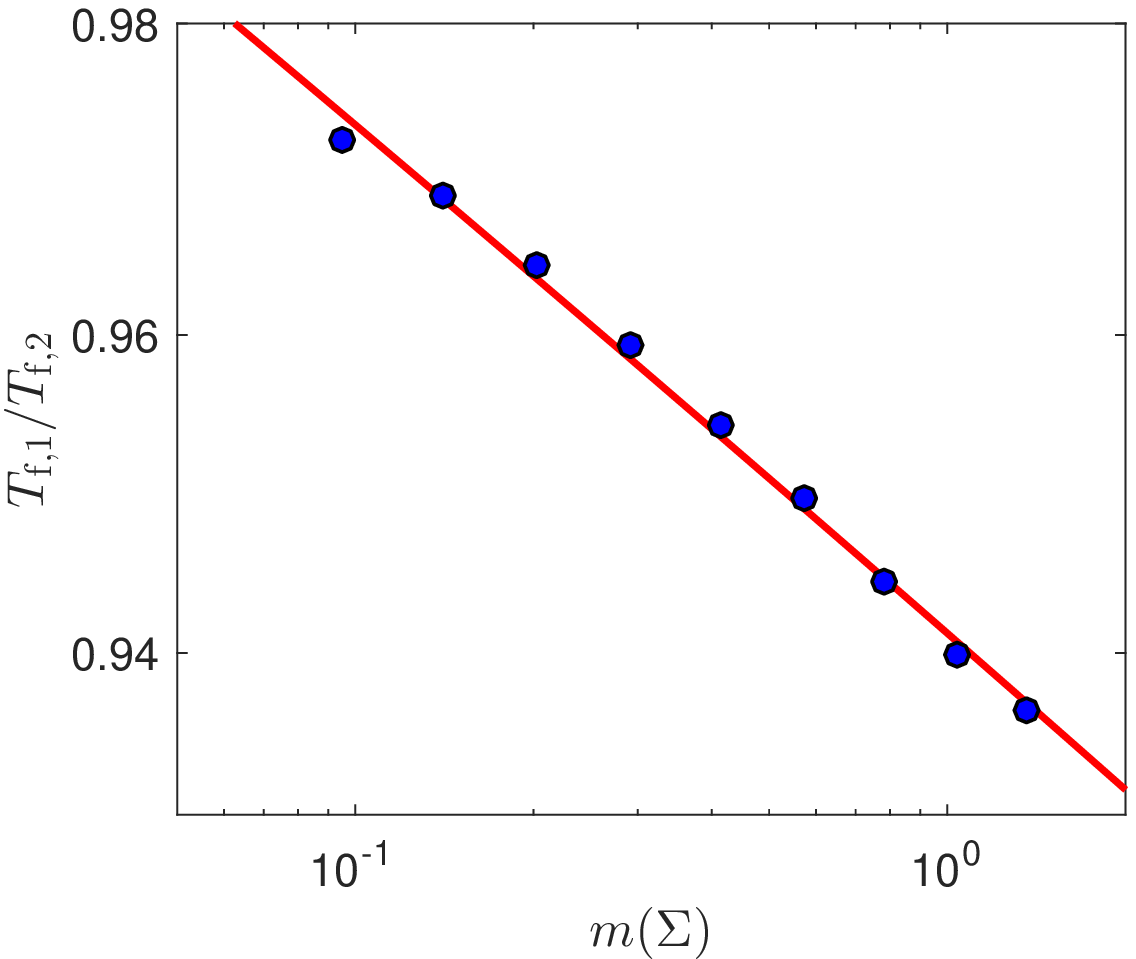}
	\end{center}
	\caption{Ratio of the fluidization times $T_{f,1}/T_{f,2}$ (symbols) predicted theoretically for the two different initial conditions used in Fig.~\ref{suppfig3}  {with $k(\tilde{f})=\tilde{f}$}. $T_{f,1}$ ($T_{f,2}$ resp.) refers to a system with the initial conditions used for the red (blue resp.) line in Fig.~\ref{suppfig3}(a). Upon changing the applied stress $\Sigma$, the ratio of the two fluidization times shows a weak dependence on $m(\Sigma)$ that is well fitted by a logarithmic dependence with slope $\simeq -0.014$ (red line).}
	\label{figvel}
\end{figure}

\end{document}